\documentstyle[titlepage,12pt]{article}
\begin{document}
\renewcommand{\theequation}{\thesection.\arabic{equation}}
\title{The Effect of Spatial Curvature on the Classical and Quantum Strings}
\author{A.L. Larsen and N. S\'{a}nchez\\
\\
\\
Observatoire de Paris,
DEMIRM. Laboratoire Associ\'{e} au CNRS \\
\hspace*{-9mm}UA 336, Observatoire de Paris et
\'{E}cole Normale Sup\'{e}rieure. \\
\hspace*{-32mm}61, Avenue de l'Observatoire, 75014 Paris, France.}
\maketitle
\begin{abstract}
We study the effects of the spatial curvature on the classical and quantum
string dynamics. We find the general solution of the circular string motion
in static Robertson-Walker spacetimes with closed or open sections. This is
given closely and completely in terms of elliptic functions. The physical
properties, string length, energy and pressure are computed and analyzed.
We find the {\it back-reaction} effect of these strings on the spacetime: the
self-consistent solution to the Einstein equations is a spatially closed
$(K>0)$ spacetime with a selected value of the curvature index $K$ (the scale
factor
is normalized to unity). No self-consistent solutions with $K\leq 0$ exist.
We semi-classically quantize the circular strings and find the mass $m$ in
each case. For $K>0,$ the very massive strings,
oscillating on the full hypersphere, have $m^2\sim K n^2\;\;(n\in N_0)$
{\it independent}
of $\alpha'$ and the level spacing {\it grows} with $n,$ while the
strings oscillating on one hemisphere (without crossing the equator) have
$m^2\alpha'\sim n$ and a {\it finite} number of states $N\sim 1/(K\alpha').$
For $K<0,$ there are infinitely many string states with masses
$m\log m\sim n,$ that is, the level spacing grows {\it slower} than $n.$

The stationary string solutions as well as the generic string fluctuations
around the center of mass are also found and analyzed in closed form.
\end{abstract}
\section{Introduction and Results}
\setcounter{equation}{0}
The propagation of strings in Friedmann-Robertson-Walker (FRW)
cosmologies has been
investigated using both exact and approximative methods, see for example
Refs.[1-8]
(as well as numerical methods, which shall not be discussed here).
Except for anti de Sitter spacetime, which has negative spatial curvature, the
cosmologies that have been considered until now, have been spatially flat. In
this paper we will consider the physical effects of a non-zero (positive or
negative) curvature index on the classical and quantum strings. The
non-vanishing components of the Riemann tensor for the generic D-dimensional
FRW line element, in comoving coordinates:
\begin{equation}
ds^2=-dt^2+a^2(t)\frac{d\vec{x} d\vec{x}}{(1+\frac{K}{4}\vec{x}\vec{x})^2},
\end{equation}
are given by:
\begin{equation}
R_{itit}=\frac{-aa_{tt}}{(1+\frac{K}{4}\vec{x}\vec{x})^2},\;\;\;\;\;\;
R_{ijij}=\frac{a^2(K+a_{t}^2)}{(1+\frac{K}{4}\vec{x}\vec{x})^4};\;\;\;\;i\neq j
\end{equation}
where $a=a(t)$ is the scale factor and $K$ is the curvature index. Clearly, a
non-zero curvature index introduces a non-zero spacetime curvature; the
exceptional case provided by $K=-a_{t}^2=\mbox{const.},$ corresponds to
the Milne-Universe. From Eqs.(1.2), it is also seen that the curvature index
has to compete with the first
derivative of the scale factor. The effects of the
curvature index are therefore conveniently discussed in the family of
FRW-universes with constant scale factor, the so-called static
Robertson-Walker spacetimes. This is the point of view we take in the
present paper.

We consider both the closed ($K>0$) and the hyperbolic ($K<0$) static
Robertson-Walker spacetimes, and all our results are compared with the
already known results in the flat ($K=0$) Minkowski spacetime. We determine the
evolution of circular strings, derive the corresponding equations of state,
discuss the question of strings as self-consistent solutions to the
Einstein equations \cite{san4}, and we perform a semi-classical
quantization. We find all the stationary string configurations in these
spacetimes and we perform a canonical quantization, using the string
perturbation series approach \cite{san1}, for a static string center of mass.
\vskip 6pt
\hspace*{-6mm}The radius of a classical circular string in the spacetime
(1.1), for $a=1,$ is determined by:
\begin{equation}
\dot{r}^2+V(r)=0;\;\;\;\;\;\;\;\;V(r)=(1-Kr^2)(r^2-b\alpha'^2),
\end{equation}
where $b$ is an integration constant. This equation is solved in terms of
elliptic functions and all solutions describe oscillating strings (Fig.1
shows the potential $V(r)$ for $K>0,\;K=0,\;K<0$).

For $K>0,$ when the spatial section is a hypersphere, the string either
oscillates on one hemisphere or on the full hypersphere. The energy is
positive while the average pressure can be positive, negative or zero;
the equation of state is given by Eqs.(3.19), (3.25).
Interestingly enough, we find that the circular strings provide a
self-consistent solution to the Einstein equations with a selected
value of the curvature
index, Eq.(3.28). Self-consistent solutions to the Einstein equations with
string sources have been found previously in the form of power law
inflationary universes \cite{san4}. We semi-classically quantize the
circular strings using the stationary phase approximation method of
Ref.\cite{das}. The
strings oscillating on one hemisphere give rise to a finite number $N_-$
of states with the following mass-formula:
\begin{equation}
m_-^2\alpha'\approx \pi\;n,\;\;\;\;\;\;\;\;N_-\approx\frac{4}{\pi K\alpha'}.
\end{equation}
As in flat Minkowski spacetime, the scale of these string states is set by
$\alpha'.$ The strings oscillating on the full hypersphere give rise to
an infinity of more and more massive states with the asymptotic mass-formula:
\begin{equation}
m_+^2\approx K\;n^2.
\end{equation}
The masses of these states are independent of $\alpha',$ the scale is set
by the curvature index $K.$ Notice also that the level spacing grows
with $n.$ A similar result was found recently for strings in anti de
Sitter spacetime \cite{san5,san6}.

For $K<0,$ when the spatial section is a hyperboloid, both the energy and
the average pressure of the oscillating strings are positive. The equation
of state is given by Eq.(3.35). In this case, the strings can not provide a
self-consistent solution to the Einstein equations. After semi-classical
quantization, we find an infinity of more and more massive states. The
mass-formula is given by Eq.(4.26):
\begin{equation}
\sqrt{-Km^2\alpha'^2}\;\log\sqrt{-Km^2\alpha'^2}\approx
-\frac{\pi}{2}K\alpha'\;n
\end{equation}
Notice that the level spacing grows faster than in
Minkowski spacetime but slower than in the closed
static Robertson-Walker spacetime.
A summary of the classical and semi-classical features of the circular
strings is presented in Tables I and II. Figs.2-4 depict the mass
quantization conditions for the $K>0$ and $K<0$ cases.
\vskip 6pt
\hspace*{-6mm}On the other hand, the stationary strings are determined by:
\begin{equation}
\phi'=\frac{L}{r^2},\;\;\;\;\;\;\;\;
r'^2+U(r)=0;\;\;\;U(r)=(1-Kr^2)(\frac{L^2}{r^2}-1),
\end{equation}
where $L$ is an integration constant (Fig.5 shows the potential
$U(r)$ for $K>0,\;K=0,\;K<0$). For $K>0,$ all the stationary string
solutions describe circular strings winding around the hypersphere.
The equation of state is of the extremely unstable string type \cite{ven}.
For $K<0,$ the stationary strings are represented by infinitely long
open configurations with an angle between the two "arms" given by:
\begin{equation}
\Delta\phi=\pi-2\arctan(\sqrt{-K}L).
\end{equation}
The energy density is positive while the pressure densities are negative.
No simple equation of state is found for these solutions.
A summary of the results for the stationary strings is presented in Table III.
\vskip 6pt
\hspace*{-6mm}Finally we compute the first and second order fluctuations
around a static string center of mass, using the string perturbation series
approach \cite{san1}
and its covariant versions \cite{san3,men}. Up to second order,
the mass-formula for arbitrary values of the curvature index
(positive or negative) is identical to the well-known flat spacetime
mass-formula; all dependence on $K$ cancels out.
\section{The Static Robertson-Walker Spacetimes}
\setcounter{equation}{0}
To clarify our notation we start by reviewing a few fundamental aspects of
the Robertson-Walker spacetimes with curved spatial sections. The general
line element is:
\begin{equation}
ds^2=-dt^2+a^2(t)[d\xi^2+f^2(\xi)d\Omega^2_{D-2}],
\end{equation}
where the function $f(\xi)$ is given by:
\begin{eqnarray}
f(\xi)\hspace*{-2mm}&=&\hspace*{-2mm}\sin\xi,\;\;\;\;0\leq\xi\leq\pi,\;\;\;\;K>0
\nonumber\\
f(\xi)\hspace*{-2mm}&=&\hspace*{-2mm}\xi,\;\;\;\;0\leq\xi<\infty,\;\;\;\;K=0\\
f(\xi)\hspace*{-2mm}&=&\hspace*{-2mm}\sinh\xi,\;\;\;\;0\leq\xi<\infty,
\;\;\;\;K<0\nonumber
\end{eqnarray}
The spatial sections are closed, flat or hyperbolic depending on whether
$K$ is positive, zero or negative. Usually a non-zero curvature index $K$
is scaled to either plus or minus $1$ by a redefinition of the scale factor
$a(t),$ but for our purposes of considering the so-called static
Robertson-Walker spacetimes, it is convenient to set the constant scale-factor
equal to unity and keep $K$ arbitrary. We shall also use the coordinates
defined by setting:
\begin{equation}
r=f(\xi),\\
\end{equation}
in which case the line element takes the
form (after a rescaling):
\begin{equation}
ds^2=-dt^2+a^2(t)[\frac{dr^2}{1-Kr^2}+r^2d\Omega^2_{D-2}],
\end{equation}
Notice that in the case of closed spatial sections, the latter coordinates
cover only half of the spatial hypersurface (for $r\in[0,1/\sqrt{K}\;]$), which
is in that case a hypersphere
of radius $1/\sqrt{K}.$ Finally it is also useful to
have the comoving coordinates defined by:
\begin{equation}
r=\frac{R}{1+\frac{K}{4}R^2},
\end{equation}
with the corresponding line element:
\begin{equation}
ds^2=-dt^2+a^2(t)[\frac{dR^2+R^2d\Omega^2_{D-2}}{(1+\frac{K}{4}R^2)^2}].
\end{equation}
For a general $D$-dimensional curved spacetime with curvature index $K,$
cosmological constant $\Lambda$ and an energy-momentum
tensor of the fluid form, the Einstein equations read:
\begin{equation}
(D-1)(D-2)(K+a_t^2)=2a^2(G\rho+\Lambda),
\end{equation}
\begin{equation}
2(D-2)aa_{tt}+(D-3)(D-2)(K+a_t^2)=2a^2(\Lambda-GP),
\end{equation}
where $\rho$ is the energy density, $P$ is the pressure, $G$ is a positive
constant (essentially the gravitational constant in $D$ dimensions)
and $a_t\equiv da(t)/dt.$ In both equations the curvature index has to
compete directly with the derivative of the scale factor, thus for instance in
inflationary models (like de Sitter or power law universes), the effect of
the curvature index will soon be negligible. In this paper we are interested
precisely
in the effects of the curvature index on the classical and quantum string
propagation, and it follows that this investigation is most suitably performed
in spacetimes with vanishing $a_t,$ and not only because of simplicity. These
spacetimes of constant scale factor are denoted the static Robertson-Walker
spacetimes. By scaling we obtain $a=1,$ and the Einstein equations take the
form:
\begin{equation}
(D-1)(D-2)K=2(G\rho+\Lambda),
\end{equation}
\begin{equation}
(D-3)(D-2)K=2(\Lambda-GP).
\end{equation}
Usually energy and pressure are
supposed to be non-negative. However, considering string sources in the
Einstein equations, this is not
necessarily true \cite{san4}, not even
in flat Minkowski spacetime \cite{san5}. In fact, in the
next section we shall return to the question of self-consistent string
solutions to the Einstein equations in the case of vanishing cosmological
constant. We will consider bosonic strings with equations of motion and
constraints given by:
\begin{eqnarray}
&\ddot{x}^\mu-x''^\mu+\Gamma^\mu_{\rho\sigma}(\dot{x}^\rho\dot{x}^\sigma-
x'^\rho x'^\sigma)=0,&\nonumber\\
&g_{\mu\nu}\dot{x}^\mu x'^\nu=g_{\mu\nu}(\dot{x}^\mu\dot{x}^\nu+x'^\mu x'^\nu)
=0,&
\end{eqnarray}
where dot and prime stand for derivative with respect to $\tau$ and
$\sigma,$ respectively. For the $2+1$ dimensional
($d\Omega^2_{D-2}=d\phi^2$) metric defined by the line element (2.4), for
$a=1,$
they take the form:
\begin{eqnarray}
\ddot{t}\hspace*{-2mm}&-&\hspace*{-2mm}t''=0,\nonumber\\
\ddot{r}\hspace*{-2mm}&-&\hspace*{-2mm}r''+\frac{Kr}{1-Kr^2}(\dot{r}^2-r'^2)-
r(1-Kr^2)(\dot{\phi^2}-\phi'^2)=0,\nonumber\\
\ddot{\phi}\hspace*{-2mm}&-&\hspace*{-2mm}\phi''+\frac{2}{r}(\dot{\phi}
\dot{r}-\phi' r')=0,\nonumber\\
\hspace*{-2mm}&-&\hspace*{-2mm}\dot{t}t'+\frac{\dot{r}r'}{1-Kr^2}
+r^2\dot{\phi}\phi'=0,\nonumber\\
\hspace*{-2mm}&-&\hspace*{-2mm}(\dot{t}^2+t'^2)+\frac{1}{1-Kr^2}(\dot{r}^2+r'^2)
+r^2(\dot{\phi}^2+\phi'^2)=0.
\end{eqnarray}
\section{Circular Strings, Physical Interpretation, Self-Consistency}
\setcounter{equation}{0}
In this section we shall give a complete description of
the evolution and physical interpretation of circular string
configurations in the static Robertson-Walker spacetimes. A plane circular
string effectively lives in $2+1$ dimensions so we will drop the dimensions
perpendicular to the string plane. The line element, in the coordinates (2.4),
is then:
\begin{equation}
ds^2=-dt^2+\frac{dr^2}{1-Kr^2}+r^2d\phi^2.
\end{equation}
Circular strings have been intensively studied in static
spacetimes \cite{san2,san3,egu,san5,mik1,mik2,all1},
but until now not in the static Robertson-Walker spacetimes. The general
equations determining the evolution and dynamics of the circular strings
\cite{san3,egu} can however be used directly here too. The ansatz
$(t=t(\tau),\;r=r(\tau),\;\phi=\sigma),$ corresponding to a
circular string, leads to \cite{san3}:
\begin{equation}
\dot{t}=\sqrt{b}\alpha',\;\;\;\;\;\;\;\;\dot{r}^2+V(r)=0,
\end{equation}
where $\alpha'$ is the string tension, $b$ is a positive
integration constant with
the dimension of $(\mbox{mass})^2$ and the potential $V(r)$ is given by:
\begin{equation}
V(r)=(1-Kr^2)(r^2-b\alpha'^2).
\end{equation}
By insertion of Eqs.(3.2)-(3.3) into Eq.(3.1), we obtain the
induced line element on the world-sheet:
\begin{equation}
ds^2=r^2(\tau)(-d\tau^2+d\sigma^2),
\end{equation}
and the string length is given by:
\begin{equation}
l(\tau)=2\pi|r(\tau)|.
\end{equation}
Energy and pressure of the circular strings can be obtained from the $2+1$
dimensional spacetime energy-momentum tensor:
\begin{equation}
\sqrt{-g}T^{\mu\nu}=\frac{1}{2\pi\alpha'}\int d\tau d\sigma
(\dot{X}^\mu\dot{X}^\nu-X'^\mu X'^\nu)\delta^{(3)}(X-X(\tau,\sigma)).
\end{equation}
After integration over a spatial volume that completely encloses the string
\cite{san4},
the energy-momentum tensor for a circular string takes the form of a
fluid:
\begin{equation}
T^\mu\;_\nu=\mbox{diag.}(-\rho,P,P),
\end{equation}
where in the comoving coordinates:
\begin{equation}
\rho=\frac{1}{\alpha'}\dot{t}=\sqrt{b},
\end{equation}
\begin{equation}
P=\frac{1}{2\alpha'\dot{t}}\;\frac{\dot{R}^2-R^2}{(1+\frac{K}{4}R^2)^2}=
\frac{1}{2\sqrt{b}\alpha'^2}[\frac{\dot{r}^2}{1-Kr^2}-r^2]=
\frac{b\alpha'^2-2r^2}{2\sqrt{b}\alpha'^2}.
\end{equation}
Let us now consider separately
the three different cases of vanishing, positive and negative curvature index.
For $K=0$ we have flat Minkowski spacetime, the circular string
configurations there
and their physical interpretation were already discussed in
Ref.\cite{san5}. The circular string potential (3.3) is shown in Fig.1a and the
equations of motion (3.2) determining the string radius are solved by:
\begin{equation}
t(\tau)=\sqrt{b}\alpha'\tau,\;\;\;\;\;r(\tau)=\sqrt{b}\alpha'\cos\tau,
\;\;\;\;\;(K=0)
\end{equation}
i.e. the string motion follows a pure harmonic motion with period
$T_{\tau}=2\pi$ in
the world-sheet time. The energy and pressure, Eqs.(3.8)-(3.9), are
given by:
\begin{equation}
\rho=\sqrt{b},
\end{equation}
\begin{equation}
P=-\frac{\sqrt{b}}{2}\cos 2\tau.
\end{equation}
Notice that during an oscilation of the string, the equation of state
"oscillates" between $P=\rho/2,$ corresponding to
{\it ultra-relativistic matter}
in $2+1$ dimensions, and $P=-\rho/2,$ corresponding to
{\it extremely unstable
strings} \cite{ven}. For further discussion on this point, see
Ref.\cite{san5}.
Using the exact time-dependent pressure (3.12), it is clear from
Eqs.(2.9)-(2.10) that the circular strings do not provide a self-consistent
solution in Minkowski spacetime. If using instead the average values over one
oscillation:
\begin{equation}
<\rho>=\sqrt{b},\;\;\;\;<P>=0,
\end{equation}
corresponding to cold matter, we see that Eq.(2.10) with vanishing
cosmological constant is fulfilled, while
Eq.(2.9) leads to:
\begin{equation}
K=G\sqrt{b}.
\end{equation}
The circular strings thus generate a positive curvature index and we conclude
that even after averaging over an oscillation, they
do not provide a self-consistent solution in Minkowski spacetime.

For positive curvature index, the spatial hypersurface is a sphere and
we have to distinguish between the two cases
$\sqrt{bK}\alpha'\leq 1$ and $\sqrt{bK}\alpha'>1,$  as is clear from the
location of the zeros of the potential, see Figs.1b,1c. For
$\sqrt{bK}\alpha'\leq 1,$ the solution of Eqs.(3.2)-(3.3) is:
\begin{equation}
t_-(\tau)=\sqrt{b}\alpha'\tau,\;\;\;\;\;r_-(\tau)=
\sqrt{b}\alpha'\mbox{sn}[\tau,k_-]\;\;\;\;\;(K>0),
\end{equation}
where $\mbox{sn}[\tau,k_-]$ is the Jacobi elliptic function and the
elliptic modulus is given by:
\begin{equation}
k_-=\sqrt{bK}\alpha'\in[0,1].
\end{equation}
The
solution describes a string oscillating between zero radius and maximal
radius $r_{\mbox{max}}=\sqrt{b}\alpha'$ with period
$T_\tau=4K(k_-)$ in the world-sheet time,
where $K(k_-)$ is the complete elliptic integral of the first
kind. Since $r_{\mbox{max}}\leq 1/\sqrt{K},$ the string oscillates on one
hemisphere; it does not cross the equator. The energy and pressure are obtained
from Eqs.(3.8)-(3.9):
\begin{equation}
\rho_-=\sqrt{b}=\frac{k_-}{\sqrt{K}\alpha'},
\end{equation}
\begin{equation}
P_-=\frac{k_-}{2\sqrt{K}\alpha'}(1-2\mbox{sn}^2[\tau,k_-]).
\end{equation}
During an oscillation of the string, the equation of state "oscillates"
between $P_-=\rho_-/2$ and $P_-=-\rho_-/2.$ This is similar to the situation
in Minkowski spacetime. The average values are given by:
\begin{equation}
<\rho_->=\frac{k_-}{\sqrt{K}\alpha'},\;\;\;\;\;<P_->
=\frac{1}{2\sqrt{K}\alpha'}[\frac{k_-^2-2}{k_-}+
\frac{2}{k_-}\;\frac{E(k_-)}{K(k_-)}],
\end{equation}
so that $2\sqrt{K}\alpha'<P_->\;\in[-1,0]$ where the limit $-1$ corresponds to
$k_-\rightarrow 1$ and the limit $0$ corresponds to $k_-\rightarrow 0.$ Thus in
average the pressure is {\it negative}
(in the limit $k_-=0,$ there are no strings
at all) and the equation of state is written:
\begin{equation}
<P_->=(\gamma(k_-)-1)<\rho_->;\;\;\;\;\;\gamma(k_-)=\frac{3}{2}-
\frac{1}{k_-^2}[1-\frac{E(k_-)}{K(k_-)}].
\end{equation}
When $k_-$
increases from $k_-=0$ to $k_-=1,$ the function $\gamma(k_-)$ decreases
from $\gamma(0)=1$ to $\gamma(1)=1/2,$ that is, from {\it cold matter} type
($\gamma=1$) to {\it extremely unstable string} type ($\gamma=1/2$).

Returning now
to the Einstein equations (2.9)-(2.10) without cosmological constant, we see
that Eq.(2.9) can be fulfilled using the average values (3.19), but Eq.(2.10)
can not. We conclude that the string solutions, Eq.(3.15), for
$\sqrt{bK}\alpha'\leq 1,$ do not provide a
self-consistent solution to the Einstein equations.

We now consider the case where
$\sqrt{bK}\alpha'>1.$ The solution of Eqs.(3.2)-(3.3) is (see Fig.1c):
\begin{equation}
t_+(\tau)=\sqrt{b}\alpha'\tau,\;\;\;\;\;r_+(\tau)=\frac{1}{\sqrt{K}}
\;\mbox{sn}[\tau/k_+,k_+],\;\;\;\;\;(K>0)
\end{equation}
where the elliptic modulus is now given by:
\begin{equation}
k_+=\frac{1}{\sqrt{bK}\alpha'}\in\;]0,1[\;.
\end{equation}
This string solution is oscillating
between zero radius and maximal radius $r_{\mbox{max}}=1/\sqrt{K},$
corresponding to the radius of the hypersphere, with period
$T_\tau=4k_+K(k_+)$ in the world-sheet
time. The physical interpretation of this solution is a
string oscillating on the full hypersphere. For $\tau=0$ it starts with zero
radius on one of the hemispheres. It expands and reaches the equator
for $\tau=k_+K(k_+).$ It then crosses the equator and contracts on the other
hemisphere until it collapses to a point for $\tau=2k_+K(k_+).$ It now expands
again, crosses the equator and eventually collapses to its initial
configuration of zero radius for $\tau=4k_+K(k_+).$
The energy and pressure of this solution are given by:
\begin{equation}
\rho_+=\sqrt{b}=\frac{1}{\sqrt{K}\alpha'k_+},
\end{equation}
\begin{equation}
P_+=\frac{1-2k_+^2\mbox{sn}^2[\tau/k_+,k_+]}{2k_+\sqrt{K}\alpha'}.
\end{equation}
During an oscillation, the equation of state "oscillates" between
$P_+=\rho_+/2$ and $P_+=(1-2k_+^2)\rho_+/2.$ The average values are given by:
\begin{equation}
<\rho_+>=\frac{1}{\sqrt{K}\alpha'k_+},\;\;\;\;\;<P_+>=
\frac{1}{2k_+\sqrt{K}\alpha'}
[\frac{2E(k_+)}{K(k_+)}-1],
\end{equation}
so that $2\sqrt{K}\alpha'<P_+>\;\in\;]-1,\infty[\;$ where the limit $-1$
corresponds to
$k_+\rightarrow 1$ and the limit $\infty$ corresponds to
$k_+\rightarrow 0.$ Thus
the average pressure can be negative, zero or positive for these solutions.
The equation of state is:
\begin{equation}
<P_+>=(\gamma(k_+)-1)<\rho_+>;\;\;\;\;\;\gamma(k_+)=
\frac{1}{2}+\frac{E(k_+)}{K(k_+)}.
\end{equation}
When $k_+$ increases from $k_+=0$ to $k_+=1,$ the function
$\gamma(k_+)$ decreases
from $\gamma(0)=3/2$ to $\gamma(1)=1/2,$ that is, from
{\it ultra-relativistic matter} type
($\gamma=3/2$) to {\it extremely unstable string} type ($\gamma=1/2$).
Let us consider also the question of self-consistency in this case. Using
the average values (3.26) in the Einstein equations (2.9)-(2.10),
without cosmological constant, we find that the self-consistency conditions
are satisfied with:
\begin{equation}
K=\frac{G}{k_+\sqrt{K}\alpha'},\;\;\;\;\;2E(k_+)=K(k_+),
\end{equation}
which yield the (numerical) solution:
\begin{equation}
k_+=0.9089...,\;\;\;\;\;K=(\frac{G}{\alpha'})^{2/3}\times 1.0658...
\end{equation}
It follows that a gas of oscillating circular strings described by
Eq.(3.21) for $k_+=0.9089...,$ provides a self-consistent solution to the
Einstein equations. The solution is a spatially closed static
Robertson-Walker spacetime with scale factor normalized to $a=1$
and curvature index
$K=(\frac{G}{\alpha'})^{2/3}\times 1.0658...$

Finally we consider circular strings in the spatially hyperbolic case,
corresponding to negative curvature index. The potential is shown in
Fig.1d. and Eqs.(3.2)-(3.3) are solved by:
\begin{equation}
t(\tau)=\sqrt{b}\alpha'\tau,\;\;\;\;\;r(\tau)=\frac{k}{\sqrt{-K}}
\;\mbox{sd}[\mu\tau,k],\;\;\;\;\;(K<0)
\end{equation}
where we introduced the notation:
\begin{equation}
\mu=\sqrt{1-Kb\alpha'^2},\;\;\;\;\;k=\sqrt{\frac{-Kb\alpha'^2}{1-Kb\alpha'^2}}
\;\in\;[0,1[\;
\end{equation}
The solution describes a string oscillating between zero radius and maximal
radius $r_{\mbox{max}}$ with period $T_\tau$ in the world-sheet time:
\begin{equation}
r_{\mbox{max}}=\frac{1}{\sqrt{-K}}\frac{k}{\sqrt{1-k^2}},\;\;\;\;\;
T_\tau=4K(k)/\mu.
\end{equation}
The energy and pressure are given by:
\begin{equation}
\rho=\sqrt{b}=\frac{k}{\sqrt{-K}\alpha'\sqrt{1-k^2}},
\end{equation}
\begin{equation}
P=\frac{k-2k(1-k^2)\mbox{sd}^2[\mu\tau,k]}{2\sqrt{1-k^2}\sqrt{-K}\alpha'}.
\end{equation}
During an oscillation, the equation of state "oscillates" between
$P=\rho/2$ and $P=-\rho/2.$ This is like the situation in flat Minkowski
spacetime. The average values are given by:
\begin{equation}
<\rho>=\frac{k}{\sqrt{-K}\alpha'\sqrt{1-k^2}},\;\;\;\;\;
<P>=\frac{1}{2\sqrt{1-k^2}\sqrt{-K}\alpha'}
[-k+\frac{2}{k}(1-\frac{E(k)}{K(k)})],
\end{equation}
so that $2\sqrt{-K}\alpha'<P>\;\in[0,\infty[\;$ where the limit $0$
corresponds to
$k\rightarrow 0$ and the limit $\infty$ corresponds to $k\rightarrow 1.$ The
average pressure is always {\it positive}
(for $k=0,$ there are no strings at all).
The equation of state takes the form:
\begin{equation}
<P>=(\gamma(k)-1)<\rho>;\;\;\;\;\;\gamma(k)=\frac{1}{2}+
\frac{1}{k^2}(1-\frac{E(k)}{K(k)}).
\end{equation}
When $k$ increases from $k=0$ to $k=1,$ the function $\gamma(k)$ increases
from $\gamma(0)=1$ to $\gamma(1)=3/2,$ that is, from {\it cold matter} type
($\gamma=1$) to {\it ultra-relativistic matter}
type ($\gamma=3/2$). Clearly, these
solutions can not provide a self-consistent solution to the Einstein
equations (2.9)-(2.10).

This concludes our analysis of classical circular string configurations
in the static Robertson-Walker spacetimes. A summary of the results
is presented in Table I.
\section{Semi-Classical Quantization}
\setcounter{equation}{0}
In this section we perform a semi-classical quantization of the circular
string configurations discussed in the previous section. We use an approach
developed in field theory by Dashen et. al. \cite{das}, based
on the stationary phase approximation of
the partition function. The method can be only used for time-periodic
solutions of
the classical equations of motion. In our string problem, these
solutions include all the circular
string solutions in the static Robertson-Walker spacetimes, discussed
in the previous section.
The method has recently been used to quantize circular strings in de
Sitter and anti de Sitter spacetimes \cite{san5} also, and we shall follow
the analysis of Ref.\cite{san5} closely.

The result of the stationary phase integration is expressed in terms of
the function:
\begin{equation}
W(m)\equiv S^{\mbox{cl}}(T(m))+m\;T(m),
\end{equation}
where $S^{\mbox{cl}}$ is the action of the classical solution, $m$ is the
mass and
the period $T(m)$ is implicitly given by:
\begin{equation}
\frac{dS^{\mbox{cl}}}{dT}=-m.
\end{equation}
In string theory we must choose $T$ to be the period in a physical
time variable. In the static Robertson-Walker spacetimes, it is
convenient to take $T$ to be the period in the comoving time $t.$ From
Eq.(3.2), it follows that:
\begin{equation}
T=\sqrt{b}\alpha' T_\tau.
\end{equation}
The bound state quantization condition is \cite{das}:
\begin{equation}
W(m)=2\pi\;n,\quad n \in N_{0},
\end{equation}
$n$ being "large". We will consider the two cases of positive and negative
curvature index. The case of zero curvature index was considered
in Ref.\cite{san5}
and the results in that case will come out anyway in the limit $K\rightarrow 0$
(from above or below).
The classical action is given by:
\begin{eqnarray}
S^{\mbox{cl}}\hspace*{-2mm}&=&\hspace*{-2mm}
\frac{1}{2\pi\alpha'}\int_0^{2\pi}d\sigma\int_0^{T_\tau}d\tau\;
g_{\mu\nu}[\dot{X}^\mu\dot{X}^\nu-X'^\mu X'^\nu]\nonumber\\
\hspace*{-2mm}&=&\hspace*{-2mm}-\frac{2}{\alpha'}\int_0^{T_\tau}d\tau\;
r^2(\tau),
\end{eqnarray}
where we used Eqs.(3.2)-(3.3).
We first consider positive $K$ in the case $\sqrt{bK}\alpha'\leq 1,$ i.e.
the solution (3.15), corresponding to strings oscillating on one
hemisphere. The period in comoving time is given by:
\begin{equation}
T_-=\frac{4k_-K(k_-)}{\sqrt{K}}.
\end{equation}
The classical action over one period becomes:
\begin{equation}
S_-^{\mbox{cl}}=\frac{8}{K\alpha'}[E(k_-)-K(k_-)].
\end{equation}
A straightforward calculation yields:
\begin{equation}
\frac{dT_-}{dk_-}=\frac{4}{\sqrt{K}}\frac{E(k_-)}{1-k_-^2},\;\;\;\;\;
\frac{dS_-^{\mbox{cl}}}{dk_-}=-\frac{8}{K\alpha'}\frac{k_-E(k_-)}{1-k_-^2}.
\end{equation}
Now Eq.(4.2) leads to:
\begin{equation}
m_-=\frac{2k_-}{\sqrt{K}\alpha'},
\end{equation}
and the quantization condition (4.4) becomes:
\begin{equation}
W_-=\frac{8}{K\alpha'}[E(k_-)-(1-k_-^2)K(k_-)]=2\pi\;n
\end{equation}
This equation determines a quantization of the parameter $k_-,$
which through Eq.(4.9)
yields a quantization of the mass. A full parametric plot of $K\alpha'W_-$
as a function of $K\alpha'^2m_-^2\;$ for $k_-\in[0,1]$ is shown in Fig.2.
A fair approximation is provided by the straight line connecting the two
endpoints:
\begin{equation}
W_-\approx 2m_-^2\alpha',
\end{equation}
so that the mass quantization condition becomes:
\begin{equation}
m_-^2\alpha'\approx \pi\;n
\end{equation}
The total number of states is then estimated to be:
\begin{equation}
N_-\approx\frac{4}{\pi K\alpha'}.
\end{equation}
This is the number of quantized circular string states oscillating on
one hemisphere without crossing the equator. The circular strings oscillating
on the full hypersphere are obtained for
$\sqrt{bK}\alpha'>1,$ and are given by the solution (3.21).
Their period in comoving time is:
\begin{equation}
T_+=\frac{4K(k_+)}{\sqrt{K}}.
\end{equation}
The classical action over one period becomes:
\begin{equation}
S_+^{\mbox{cl}}=\frac{8}{K\alpha'}\;\frac{E(k_+)-K(k_+)}{k_+}.
\end{equation}
A straightforward calculation yields:
\begin{equation}
\frac{dT_+}{dk_+}=\frac{4}{\sqrt{K}}[\frac{E(k_+)}{k_+(1-k_+^2)}-
\frac{K(k_+)}{k_+}],
\;\;\;\;\;\frac{dS_+^{\mbox{cl}}}{dk_+}=-\frac{8}{K\alpha'}[
\frac{E(k_+)}{k_+^2(1-k_+^2)}-\frac{K(k_+)}{k_+^2}].
\end{equation}
Now Eq.(4.2) leads to:
\begin{equation}
m_+=\frac{2}{\sqrt{K}\alpha'k_+},
\end{equation}
and the quantization condition (4.4) becomes:
\begin{equation}
W_+=\frac{8}{K\alpha'}\frac{E(k_+)}{k_+}=2\pi\;n
\end{equation}
This equation determines a quantization of the parameter $k_+,$ which
through Eq.(4.17)
yields a quantization of the mass. A parametric plot of $K\alpha'W_+$
as a function of $K\alpha'^2m_+^2\;$ for $k_+\in\;]0,1[\;$ is shown in Fig.3.
Asymptotically, when the mass grows indefinetely, corresponding to
$k_+\rightarrow 0$ (see Eq.(4.17)), we find:
\begin{equation}
K\alpha'^2m_+^2\approx\frac{4}{k_+^2},\;\;\;\;\;K\alpha'W_+
\approx\frac{4\pi}{k_+},
\end{equation}
and the mass quantization condition becomes:
\begin{equation}
m_+^2\approx K\;n^2.
\end{equation}
Several interesting remarks are now in order: First, notice that the mass is
independent of $\alpha'.$ The {\it scale}
of these very massive states is set by
the curvature index $K.$ Secondly, since the mass is proportional to $n,$
the level spacing ($\Delta(m^2\alpha')$ as a function of $n$) {\it grows}
proportionally to $n.$ This is completely different from flat
Minkowski spacetime where the level spacing is constant. Finally, it should
be mentioned that the same behaviour for very massive strings was
found recently in anti de Sitter spacetime \cite{san5,san6}. At
this point it is
tempting to consider the partition function for a gas of strings at finite
temperature, but it must be stressed that a discussion of the thermodynamic
properties (for instance existence or non-existence of a Hagedorn
temperature) must be based on exact quantization of generic strings,
and not only on a semi-classical
quantization of special circular string configurations. It must be noticed,
however, that it has been shown recently \cite{san5,san6}, that for Minkowski,
de Sitter and anti de Sitter spacetimes, the spectrum of generic strings and
the semi-classical spectrum of circular strings are in complete agreement.

Let us now consider the semi-classical quantization of the circular strings
in the spatially hyperbolic spacetime, i.e. we return to the solutions (3.29).
The period in comoving time is:
\begin{equation}
T=\frac{4kK(k)}{\sqrt{-K}}.
\end{equation}
The classical action over one period becomes:
\begin{equation}
S^{\mbox{cl}}=\frac{8}{K\alpha'}\;\frac{E(k)-(1-k^2)K(k)}{\sqrt{1-k^2}}.
\end{equation}
A straightforward calculation yields:
\begin{equation}
\frac{dT}{dk}=\frac{4}{\sqrt{-K}}\;\frac{E(k)}{1-k^2},
\;\;\;\;\;\frac{dS^{\mbox{cl}}}{dk}=\frac{8}{K\alpha'}\;
\frac{kE(k)}{(1-k^2)^{3/2}}.
\end{equation}
Now Eq.(4.2) leads to:
\begin{equation}
m=\frac{2}{\sqrt{-K}\alpha'}\;\frac{k}{\sqrt{1-k^2}},
\end{equation}
and the quantization condition (4.4) becomes:
\begin{equation}
W=\frac{8}{K\alpha'}\frac{E(k)-K(k)}{\sqrt{1-k^2}}=2\pi\;n
\end{equation}
This equation determines a quantization of the parameter $k,$ which
through Eq.(4.24)
yields a quantization of the mass. A parametric plot of $\;-K\alpha'W$
as a function of $\;-K\alpha'^2m^2\;$ for $k\in[0,1[\;$ is shown in Fig.4.
Asymptotically, when the mass grows indefinetely, corresponding to
$k\rightarrow 1$ (see Eq.(4.24)), we find the quantization condition:
\begin{equation}
\sqrt{-Km^2\alpha'^2}\;\log\sqrt{-Km^2\alpha'^2}\approx
-\frac{\pi}{2}K\alpha'\;n
\end{equation}
Formally, it corresponds to a mass-formula in the form
$"m\log m\propto n"$, i.e. the level spacing grows faster than in
Minkowski spacetime (where it is constant) but slower than in the closed
static Robertson-Walker spacetime (where it grows proportional to $n$).

We close this section by ensuring that our results reproduce correctly the
spectrum in flat Minkowski spacetime by taking the limit $K\rightarrow 0.$
Starting for instance from Eqs.(4.9)-(4.10) in the spatially closed
universe, we find in the limit $K\rightarrow 0:$
\begin{equation}
m=2\sqrt{b},\;\;\;\;\;W=2\pi\sqrt{b}\alpha'
\end{equation}
which by Eq.(4.4) gives:
\begin{equation}
m^2\alpha'=4\;n
\end{equation}
If we subtract the intercept $-4,$ this is the well-known (exact)
mass formula for closed bosonic strings in flat Minkowski spacetime. Same
result is obtained by starting from Eqs.(4.17)-(4.18) or from
Eqs.(4.24)-(4.25).

The main conclusions of this section are presented in Table II.
\section{Stationary Strings}
\setcounter{equation}{0}
In this section we will supplement our results on exact classical string
solutions by considering the family of stationary strings. Stationary
strings certainly do not tell much about string propagation in curved
spacetimes, which is our main aim here, but since they are equilibrium
configurations, existing only when there is an exact balance between
the local gravity and the string tension, their actual shape reflects the
geometry and topology of the underlying spacetime and they provide
information about the interaction of gravity on strings. We shall follow a
recent approach \cite{all2} where the stationary strings are described by a
potential in the (stationary) radial coordinate.
The stationary string ansatz $(t=\tau,\;r=r(\sigma),\;\phi=\phi(\sigma))$
in Eq.(2.12) leads to:
\begin{equation}
\phi'=\frac{L}{r^2},\;\;\;\;\;r'^2+U(r)=0,
\end{equation}
where $L$ is an integration constant and the potential $U(r)$ is given by:
\begin{equation}
U(r)=(1-Kr^2)(\frac{L^2}{r^2}-1).
\end{equation}
Notice that we consider plane
stationary strings in the backgrounds (3.1), for $K$ positive, negative or
zero.
Possible extra transverse dimensions have been dropped for simplicity.
Beside energy and pressure, it is also interesting to consider the string
length. By insertion of the stationary string ansatz and Eqs.(5.1)-(5.2)
in the line element (3.1),
we find that the string length element is:
\begin{equation}
dl=d\sigma,
\end{equation}
thus $\sigma$ measures directly the length of stationary strings in the static
Robertson-Walker spacetimes.

For vanishing curvature index, corresponding to flat Minkowski spacetime, it is
well-known that the only stationary strings are the straight ones. Indeed, the
potential $U(r)$ is given by (Fig.5a.):
\begin{equation}
U(r)=\frac{L^2}{r^2}-1,
\end{equation}
and Eqs.(5.1) are solved by:
\begin{equation}
r(\sigma)=\sqrt{\sigma^2+L^2},
\end{equation}
\begin{equation}
\phi(\sigma)=\arctan(\sigma/L),
\end{equation}
which for $\sigma\in\;]-\infty,\;+\infty[\;$ describes an infinitely long
straight string parallel to the $y$-axis with "impact-parameter" $L.$
The string energy and pressure densities are obtained from Eq.(3.6):
\begin{equation}
\frac{d\rho}{dl}=\frac{d\rho}{d\sigma}=
\frac{d}{d\sigma}\int d^3 X \sqrt{-g}T^{00}=
\frac{1}{2\pi\alpha'},
\end{equation}
\begin{equation}
\frac{dP_y}{dl}=\frac{dP_y}{d\sigma}=
\frac{d}{d\sigma}\int d^3 X \sqrt{-g}T^{y}\;_{y}=
-\frac{1}{2\pi\alpha'},
\end{equation}
while $P_x=0.$ Obviously the integrated energy and pressure are infinite.
Eqs.(5.7)-(5.8) represent the well-known string equation of state in $1+1$
effective dimensions.

Let us now turn to the more interesting cases of non-vanishing curvature index.
For positive $K,$ the potential is given by (Fig.5b.):
\begin{equation}
U(r)=(1-|K|r^2)(\frac{L^2}{r^2}-1),
\end{equation}
For $L>1/\sqrt{K}$ there are no solutions since the "impact-parameter" $L$
is larger
than the radius of the hypersphere, thus we need only consider the case
$L\leq 1/\sqrt{K}.$ The solution of Eqs.(5.1) is:
\begin{equation}
r^2(\sigma)=\frac{1}{K}\;\frac{KL^2+\tan^2(\sqrt{K}\sigma)}
{1+\tan^2(\sqrt{K}\sigma)},
\end{equation}
\begin{equation}
\phi(\sigma)=\arctan[\frac{\tan(\sqrt{K}\sigma)}{\sqrt{K}L}].
\end{equation}
Consider first the "degenerate" cases $L=0$ and $L=1/\sqrt{K}:$
\begin{equation}
\phi=\pi/2,\;\;\;\;\;r(\sigma)=\pm\frac{1}{\sqrt{K}}\cos
(\sqrt{K}\sigma)\;;\;\;\;\;\;\;\;\;
\mbox{for}\;\;\;\;\;L=0
\end{equation}
\begin{equation}
\phi=\sqrt{K}\sigma,\;\;\;\;\;r(\sigma)=1/\sqrt{K}\;;\;\;\;\;\;\;\;\;
\mbox{for}\;\;\;\;\;L=1/\sqrt{K}
\end{equation}
Eqs.(5.12) describe a stationary circular string winding around the hypersphere
from pole to pole, while Eqs.(5.13) describe a stationary circular
string winding around the hypersphere along the equator. More generally,
the solution (5.10)-(5.11) describes a stationary circular string of radius
$1/\sqrt{K}$ winding around the hypersphere
for arbitrary values of $L,$ as can be seen as follows: The hypersphere is
parametrized by:
\begin{eqnarray}
x\hspace*{-2mm}&=&\hspace*{-2mm}\frac{1}{\sqrt{K}}\sin\xi\sin\phi\nonumber\\
y\hspace*{-2mm}&=&\hspace*{-2mm}\frac{1}{\sqrt{K}}\cos\xi\\
z\hspace*{-2mm}&=&\hspace*{-2mm}\frac{1}{\sqrt{K}}\sin\xi\cos\phi\nonumber
\end{eqnarray}
where $\sin\xi=\sqrt{K}r.$
A rotation by the angle $\theta$ in the $y-z\;$ plane leads to:
\begin{eqnarray}
\tilde{x}\hspace*{-2mm}&=&\hspace*{-2mm}x\nonumber\\
\tilde{y}\hspace*{-2mm}&=&\hspace*{-2mm}y\cos\theta-z\sin\theta\\
\tilde{z}\hspace*{-2mm}&=&\hspace*{-2mm}y\sin\theta+z\cos\theta\nonumber
\end{eqnarray}
Consider now the general solution (5.10)-(5.11)
and take $\theta=\arccos(\sqrt{K}L):$
\begin{eqnarray}
\tilde{x}\hspace*{-2mm}&=&\hspace*{-2mm}\frac{1}{\sqrt{K}}
\sin(\sqrt{K}\sigma)\nonumber\\
\tilde{y}\hspace*{-2mm}&=&\hspace*{-2mm}0\\
\tilde{z}\hspace*{-2mm}&=&\hspace*{-2mm}\frac{1}{\sqrt{K}}
\cos(\sqrt{K}\sigma)\nonumber
\end{eqnarray}
i.e. the solution (5.13). All stationary strings are identical up to rotations,
so we need only consider (say) the solution (5.13). By integrating the
components of the energy-momentum tensor, Eq.(3.6), we find:
\begin{equation}
\rho=\frac{1}{\sqrt{K}\alpha'},
\end{equation}
\begin{equation}
P_x=P_z=-\frac{1}{2\sqrt{K}\alpha'},
\end{equation}
corresponding to the {\it extremely
unstable string} type equation of state $P=-\rho/2$
in $2+1$ effective dimensions.

For negative curvature index, the potential (5.2) is:
\begin{equation}
U(r)=(1+|K|r^2)(\frac{L^2}{r^2}-1),
\end{equation}
and stationary strings can only exist for $r\geq L,$ see Fig.5c. Eqs.(5.1)
are solved by:
\begin{equation}
r^2(\sigma)=\frac{1}{K}\;\frac{KL^2-\tanh^2(\sqrt{-K}\sigma)}
{1-\tanh^2(\sqrt{-K}\sigma)},
\end{equation}
\begin{equation}
\phi(\sigma)=\pm\left\{\frac{\pi}{2}-\arctan[\sqrt{-K}L\tanh^{-1}
(\sqrt{-K}\sigma)]\right\}.
\end{equation}
For $\sigma\in\;]-\infty,\;+\infty[\;,$ the solution describes a string
stretching from spatial infinity towards $r=L$ and back towards spatial
infinity. The angle between the two "arms" is given by:
\begin{equation}
\Delta\phi=\pi-2\arctan(\sqrt{-K}L)\;\;\in\;\;]0,\pi].
\end{equation}
The total string length and energy are infinite, while
the energy density is given by:
\begin{equation}
\frac{d\rho}{d\l}=\frac{1}{2\pi\alpha'}.
\end{equation}
The pressures in the two directions are generally different due to lack of
symmetry. In the comoving coordinates (2.6) we find the following integral
expressions, using also Eq.(3.6):
\begin{equation}
P_x=-\frac{1}{2\pi\alpha'}\int_{-\infty}^{+\infty}d\sigma\;
[R'\cos\phi-R\phi'\sin\phi]^2,
\end{equation}
\begin{equation}
P_y=-\frac{1}{2\pi\alpha'}\int_{-\infty}^{+\infty}d\sigma\;
[R'\sin\phi-R\phi'\cos\phi]^2,
\end{equation}
where $R$ and $r$ are related by Eq.(2.5). Using the explicit solutions,
Eqs.(5.20)-(5.21), the pressure densities are:
\begin{equation}
\frac{dP_x}{dl}=\frac{dP_x}{d\sigma}=\frac{8KL^2}{\pi\alpha'(1-KL^2)^2}
e^{-\sqrt{-K}\mid\sigma\mid}
\end{equation}
\begin{equation}
\frac{dP_y}{dl}=\frac{dP_y}{d\sigma}=\frac{-8}{\pi\alpha'(1-KL^2)^2}
e^{-\sqrt{-K}\mid\sigma\mid}
\end{equation}
The pressure densities are negative ($K$ is negative) but asymptotically
go to zero exponentially. The integrated pressures are thus {\it finite}. In
fact, Eqs.(5.24)-(5.25) can be integrated explicitly:
\begin{equation}
P_x=\frac{2}{\pi\alpha'\sqrt{-K}(1-KL^2)}\left\{-\frac{2}{3}+\frac{1}{KL^2}-
\frac{1-KL^2}{KL^3\sqrt{-K}}[\frac{\pi}{2}-\arcsin\frac{1}{\sqrt{1-KL^2}}]
\right\},
\end{equation}
\begin{equation}
P_y-P_x=\frac{-4}{3\pi\alpha'\sqrt{-K}(1-KL^2)}\;<\;0
\end{equation}
In the two extreme limits $L\rightarrow\infty,\;L\rightarrow 0,$ we have:
\begin{equation}
P_x\rightarrow 0,\;\;\;\;\;P_y\rightarrow 0\;;\;\;\;\;\;\;\;\;
\mbox{for}\;\;\;\;\;L\rightarrow\infty
\end{equation}
\begin{equation}
P_x\rightarrow 0,\;\;\;\;\;P_y\rightarrow \frac{-4}{3\pi\alpha'\sqrt{-K}}\;;
\;\;\;\;\;\;\;\;\mbox{for}\;\;\;\;\;L\rightarrow 0
\end{equation}
where the latter case describes a straight string through $r=0.$
This concludes our discussion of the stationary strings and their physical
interpretation in the static Robertson-Walker spacetimes.
A summary of the results is presented in Table III.
\section{String Perturbation Series Approach}
\setcounter{equation}{0}
Until now, our analysis has been based on exact solutions to the string
equations of motion and constraints (2.11). However, it
is interesting to consider also approximative methods. In this section
we use the string perturbation series approach, originally developed
by de Vega and S\'{a}nchez \cite{san1}, to describe
string fluctuations around
the string center of mass. The target space coordinates are expanded as:
\begin{equation}
x^\mu(\tau,\sigma)=q^\mu(\tau)+\eta^\mu(\tau,\sigma)+\xi^\mu(\tau,\sigma)+...
\end{equation}
and after insertion of this series into Eqs.(2.12),
one solves the string equations of motion and constraints order
by order in the expansion. For a massive string, the zeroth order equations
determining the string center of mass read:
\begin{equation}
\ddot{q}^\mu+\Gamma^\mu_{\rho\sigma}\dot{q}^\rho\dot{q}^\sigma=0,\;\;\;\;\;
g_{\mu\nu}\dot{q}^\mu\dot{q}^\nu=-m^2\alpha'^2
\end{equation}
For a string with center of mass in the (say) $x-y$ plane of the
$D$-dimensional static Robertson-Walker spacetime (2.4), Eqs.(6.2) lead to:
\begin{eqnarray}
t\hspace*{-2mm}&=&\hspace*{-2mm}\sqrt{p^2+m^2}\;\alpha'\tau,\nonumber\\
\dot{\phi}\hspace*{-2mm}&=&\hspace*{-2mm}\frac{L\alpha'}{r^2},\\
\dot{r}^2\hspace*{-2mm}&=&\hspace*{-2mm}(1-Kr^2)(p^2-\frac{L^2}{r^2})
\alpha'^2,\nonumber
\end{eqnarray}
where $p$ and $L$ are integration constants with the physical interpretation
of momentum and angular momentum, respectively. Eqs.(6.3) can be easily solved
in terms of elementary functions, but we shall not need the general solutions
here. The
equations for the first and second order string fluctuations will turn
out to be quite complicated in the general case, so we shall consider here
only a static string center of mass, $p=L=0:$
\begin{equation}
t=m\alpha'\tau,\;\;\;\;\;r=\mbox{const.}\equiv r_0,\;\;\;\;\;\mbox{all angular
coordinates constant.}
\end{equation}
It is convenient to consider from the beginning only first order string
fluctuations in the directions perpendicular to the geodesic of the center
of mass. We thus introduce normal vectors $n^\mu_R,\;\;R=1,2,...,(D-1)$:
\begin{equation}
\eta^\mu=\delta x^R n^\mu_R,
\end{equation}
where $\delta x^R$ are the comoving fluctuations, i.e. the fluctuations
as seen by an observer travelling with the center of mass of the string. It
can be shown that the first order fluctuations fulfill the equations
\cite{san3}:
\begin{equation}
\ddot{C}_{nR}+(n^2\delta_{RS}-
R_{\mu\rho\sigma\nu}n^\mu_R n^\nu_S\dot{q}^\rho\dot{q}^\sigma)C^S_n=0,
\end{equation}
where $R_{\mu\rho\sigma\nu}$ is the Riemann tensor of the background and
$C^R_n$ are the modes of the fluctuations:
\begin{equation}
\delta x_R(\tau,\sigma)=\sum_n C_{nR}(\tau)e^{-in\sigma}
\end{equation}
In the present case of static Robertson-Walker spacetimes,
the Riemann tensor is non-zero but the projections appearing in Eq.(6.6)
actually vanish, as can be easily verified. It follows from the
explicit expressions of the normal vectors:
\begin{equation}
n^r=(0,\;\sqrt{1-Kr_0^2},\;0),\;\;\;\;\;n^{i}=(0,0,...,0,\frac{1}{r_0},0,...,0,0),
\end{equation}
that the first
order string fluctuations are ordinary plane waves:
\begin{equation}
\eta^{t}(\tau,\sigma)=0,
\end{equation}
\begin{equation}
\eta^r(\tau,\sigma)=\sqrt{1-Kr_0^2}\;\sum_n [A_ne^{-in(\sigma+\tau)}+
\tilde{A}_ne^{-in(\sigma-\tau)}],
\end{equation}
\begin{equation}
\eta^{i}(\tau,\sigma)=\frac{1}{r_0}\sum_n [A^{i}_n e^{-in(\sigma+\tau)}+
\tilde{A}^{i}_ne^{-in(\sigma-\tau)}].
\end{equation}
Here $\eta^{t},$ $\eta^r$ and $\eta^{i}$ denote the fluctuations
in the temporal, radial and
angular directions, respectively. The second order fluctuations are determined
by:
\begin{equation}
\ddot{\xi}^{t}-\xi''^t=0,
\end{equation}
\begin{equation}
\ddot{\xi^r}-\xi''^r=-\frac{Kr_0}{1-Kr_0^2}[(\dot{\eta}^r)^2-
(\eta'^r)^2]+r_0(1-Kr_0^2)\sum_{i}[(\dot{\eta}^{i})^2-(\eta'^{i})^2],
\end{equation}
\begin{equation}
\ddot{\xi}^{i}-\xi''^{i}=-\frac{2}{r_0}\sum_{i}[\dot{\eta}^r\dot{\eta}^{i}-
\eta'^r \eta'^{i}].
\end{equation}
These are just ordinary wave equations with source terms, and can be easily
solved. Thereafter, we have to expand also the constraint equations. In the
present case we find up to second order:
\begin{equation}
m^2\alpha'^2=\frac{1}{1-Kr_0^2}[(\dot{\eta}^r)^2+(\eta'^r)^2]+
r_0^2\sum_{i}[(\dot{\eta}^{i})^2+(\eta'^{i})^2],
\end{equation}
\begin{equation}
\frac{1}{1-Kr_0^2}\dot{\eta}^r \eta'^r+r_0^2\sum_{i}\dot{\eta}^{i}
\eta'^{i}=0.
\end{equation}
Introducing the notation $A_n=A^0_n,\;\;A^\alpha_n=(A^0_n,\;A^{i}_n),\;$
these equations read explicitly:
\begin{equation}
m^2\alpha'^2=-2\sum_\alpha\sum_{n,l}
l(n-l)[A^\alpha_{n-l}A^\alpha_l e^{-in(\sigma+\tau)}+\tilde{A}^\alpha_{n-l}
\tilde{A}^\alpha_l e^{-in(\sigma-\tau)}],
\end{equation}
\begin{equation}
\sum_\alpha\sum_{n,l}[A^\alpha_{n-l}A^\alpha_l
e^{-in(\sigma+\tau)}-\tilde{A}^\alpha_{n-l}
\tilde{A}^\alpha_l e^{-in(\sigma-\tau)}]=0,
\end{equation}
which are just the usual flat spacetime constraints. All dependence of the
curvature index $K$ (positive or negative) has canceled out in these formulae.
For $n=0,$ in particular, we get the usual flat spacetime mass-formula:
\begin{equation}
m^2\alpha'^2=2\sum_\alpha\sum_{l}
l^2[A^\alpha_{l}A^\alpha_{-l}+\tilde{A}^\alpha_{l}
\tilde{A}^\alpha_{-l}],
\end{equation}
with the constraint that there must be an equal amount of left and right
movers:
\begin{equation}
\sum_\alpha\sum_{l}
l^2[A^\alpha_{l}A^\alpha_{-l}-\tilde{A}^\alpha_{l}
\tilde{A}^\alpha_{-l}]=0.
\end{equation}
Notice that the spectrum found here (the flat spacetime spectrum) is very
different (when $K\neq 0$) from the spectrum of the circular strings, as
discussed in Section 4. This is however in no ways contradictory. The
circular string ansatz merely picks out particular states in the complete
spectrum, so there is no apriory reason to believe that the circular
string spectrum should be similar to the generic spectrum. Furthermore,
the perturbation approach used in this section is also semi-classical
in nature, in the sense that it is based on fluctuations around a special
solution, namely the static string center of mass, Eq.(6.4).
\section{Concluding Remarks}
We have solved the equations of motion and constraints for circular strings
in static Robertson-Walker spacetimes. We computed the equations of state
and found a self-consistent solution to the Einstein equations. The solutions
have been quantized semi-classically using the stationary phase approximation
method and the resulting spectra were analyzed and discussed. We also found all
stationary string configurations in these spacetimes and we computed the
corresponding physical quantities, string length, energy and pressure.
Finally we calculated the first and second order string fluctuations
around a static center of mass, using the string perturbation series approach.
\vskip 48pt
\hspace*{-6mm}{\bf Acknowledgements:}\\
A.L. Larsen is supported by the Danish Natural Science Research
Council under grant No. 11-1231-1SE

\newpage
\begin{centerline}
{\bf Figure Captions}
\end{centerline}
\vskip 24pt
\hspace*{-6mm}Fig.1. The potential $V(r)$ introduced in Eqs.(3.2)-(3.3)
for a circular string in the static Robertson-Walker spacetimes:
(a) flat $(K=0),$ (b) closed $(K>0\;\;\mbox{and}\;\;\sqrt{bK}\alpha'\leq 1),$
(c) closed $(K>0\;\;\mbox{and}\;\;\sqrt{bK}\alpha'>1),$
(d) hyperbolic $(K<0).$ In the cases of closed spatial
sections, the radial coordinate is only defined up to the equator
$r_{\mbox{max}}=1/\sqrt{K}\;$ (=3, in the cases shown).
\vskip 12pt
\hspace*{-6mm}Fig.2. Parametric plot of $K\alpha'W_-$ as a function of
$K\alpha'^2m_-^2,$ Eqs.(4.9)-(4.10), for $k\in[0,1]$ in the closed
static Robertson-Walker spacetime for $\sqrt{bK}\alpha'\leq 1.$
Notice that $K\alpha'W_-\in[0,8]$ and
$K\alpha'^2m_-^2\in[0,4].$ For $W_-=2\pi n\;(n\geq 0)$ there can only
be a {\it finite} number of states.
\vskip 12pt
\hspace*{-6mm}Fig.3. Parametric plot of $K\alpha'W_+$ as a function of
$K\alpha'^2m_+^2,$ Eqs.(4.17)-(4.18), for $k\in\;]0,1[\;$ in the closed
static Robertson-Walker spacetime for $\sqrt{bK}\alpha'>1.$
Notice that $K\alpha'W_+\in\;]8,\infty[\;$ and
$K\alpha'^2m_+^2\in\;]4,\infty[\;.$ For $W_+=2\pi n\;(n\geq 0)$ there will
be infinitely many states.
\vskip 12pt
\hspace*{-6mm}Fig.4. Parametric plot of $-K\alpha'W$ as a function of
$-K\alpha'^2m^2,$ Eqs.(4.24)-(4.25), for $k\in[0,1[\;$ in the hyperbolic
static Robertson-Walker spacetime.
Notice that $-K\alpha'W\in[0,\infty[\;$ and
$-K\alpha'^2m^2\in[0,\infty[\;.$ For $W=2\pi n\;(n\geq 0)$ there will
be infinitely many states.
\vskip 12pt
\hspace*{-6mm}Fig.5. The potential $U(r)$ introduced in Eqs.(5.1)-(5.2)
for a stationary string in the static Robertson-Walker spacetimes:
(a) flat $(K=0),$ (b) closed $(K>0\;\;\mbox{and}\;\;\sqrt{K}L\leq 1),$
(c) hyperbolic $(K<0).$
In the case of closed spatial
sections, the radial coordinate is only defined up to the equator
$r_{\mbox{max}}=1/\sqrt{K}\;$ (=4, in the case shown).
\newpage
\begin{centerline}
{\bf Table Captions}
\end{centerline}
\vskip 24pt
\hspace*{-6mm}Table I. Classical circular strings in the static
Robertson-Walker spacetimes. Notice that a self-consistent solution to the
Einstein equations, with the string back-reaction included, can be obtained
only for $K>0.$
\vskip 12pt
\hspace*{-6mm}Table II. Semi-classical quantization of the circular strings
in the static Robertson-Walker spacetimes. Notice in particular the
different behaviour of the high mass spectrum of strings in the three cases.
\vskip 12pt
\hspace*{-6mm}Table III. Stationary strings in the static
Robertson-Walker spacetimes. Notice that the pressure densities are
always negative in all three cases.

\newpage
\begin{thebibliography}{11}
\bibitem{san1}H.J. de Vega and N. S\'{a}nchez, Phys. Lett.
              {\bf B197} (1987) 320.
\bibitem{san2}H.J. de Vega, A.L. Larsen and N. S\'anchez,
              Nucl. Phys. {\bf B427}, 643 (1994).
\bibitem{san3}A.L. Larsen and N. S\'anchez, ``Strings propagating in the
             $2+1$ black hole anti-de Sitter spacetime'', DEMIRM 94013
preprint,
             to be published in Phys. Rev. {\bf D}.
\bibitem{ven}N. S\'anchez and G. Veneziano, Nucl. Phys.
             {\bf B333} (1990) 253.\\
             M. Gasperini, N. S\'anchez and G. Veneziano,\\
             IJMP {\bf A6} (1991) 3853, Nucl. Phys. {\bf B364} (1991) 365.
\bibitem{egu}H.J. de Vega and I.L. Egusquiza,
             Phys. Rev. {\bf D49}, 763 (1994).
\bibitem{san4}H.J. de Vega and N. S\'anchez, ``Back reaction of strings in
             self consistent string cosmology'',
             LPTHE 94-21/Demirm 94014, to
             be published in Phys. Rev. {\bf D}.
\bibitem{san5}H.J. de Vega, A.L. Larsen and N. S\'anchez,
              "Semi-Classical Quantization of Circular Strings in de Sitter and
              anti de Sitter Spacetimes", Demirm 94049,
              submitted to Phys. Rev. {\bf D}.
\bibitem{san6}A.L. Larsen and N. S\'anchez, ``Mass spectrum of strings in
              anti de Sitter spacetime'', DEMIRM 94048,
              submitted to Phys. Rev. {\bf D}.
\bibitem{das}R. Dashen, B. Hasslacher and A. Neveu, Phys. Rev. {\bf D11},
             3424 (1975).
\bibitem{men}P.F. Mende in
             "String Quantum Gravity and the Physics at the
              Planck Scale", Proceedings of the Erice Workshop held in June
              1992. Edited by N. S\'{a}nchez, World Scientific, 1993.
              Pages 286-299.
\bibitem{mik1}H.J. de Vega, A.V. Mikhailov and N. S\'{a}nchez, \\
              Teor. Mat. Fiz. {\bf 94} (1993) 232.
\bibitem{mik2}F. Combes, H.J. de Vega, A.V. Mikhailov and N. S\'{a}nchez,\\
              Phys. Rev. {\bf D50}, 2754 (1994).
\bibitem{all1}A.L. Larsen, Phys. Rev. {\bf D50} (1994) 2623.
\bibitem{all2}A.L. Larsen and N. S\'anchez, "New Classes of Exact
              Multi-String Solutions in Curved Spacetimes",
              DEMIRM preprint, January 1995.
\end{thebibliography}
\end{document}